\documentclass[twocolumn,prc,amssymb, aps,superscriptaddress, 
amsmath]{revtex4}

\usepackage{color}
\usepackage{amsmath,mathtools,booktabs,
            microtype}

\usepackage{graphicx}
\usepackage{dcolumn}
\usepackage{bm}
\usepackage{multirow}
\usepackage{times}
\usepackage{url}
\usepackage{tikz}
\usepackage[version=4]{mhchem}

\usetikzlibrary{arrows,decorations.pathmorphing}
\usetikzlibrary{arrows.meta}
\usetikzlibrary{calc}

\definecolor{violet}{HTML}{602969}
\definecolor{red}{HTML}{FC0009}
\definecolor{orange}{HTML}{FF6319}
\definecolor{green}{HTML}{00933C}
\definecolor{blue}{HTML}{0036A6}
\definecolor{yellow}{HTML}{FFBE00}
\definecolor{lightgrey}{HTML}{A7A9AC}

\newcommand{\be}{\begin{equation}}
\newcommand{\ee}{  \end{equation}}
\newcommand{\ba}{\begin{eqnarray}}
\newcommand{\ea}{  \end{eqnarray}}

\begin{document}

\title{The Rise of Stochasticity in Physics}

\author{Hans A. \surname{Weidenm\"uller}}
\email{haw@mpi-hd.mpg.de}
\affiliation{Max-Planck-Institut f\"ur Kernphysik, Saupfercheckweg 1, D-69117 Heidelberg, Germany}

\date{\today}

\begin{abstract}In the last 175 years, the physical understanding of
  nature has seen a revolutionary change. Until about 1850, Newton's
  theory and the mechanical world view derived from it provided the
  dominant view of the physical world, later supplemented by Maxwell's
  theory of the electromagnetic field. That approach was entirey
  deterministic and free of probabilistic concepts. In contrast to
  that conceptual edifice, today many fields of physics are governed
  by probabilistic concepts. Statistical mechanics in its classical or
  quantum version and random-matrix theory are obvious
  examples. Quantum mechanics is an intrinsically statistical
  theory. Classical chaos and its quantum manifestations also require
  a stochastic approach.

  The paper desribes how a combination of discoveries and conceptual
  problems undermined the mechanical world view, led to novel
  concepts, and shaped the modern understanding of physics. As
  essential causes I list Clausius' formulation of the second law of
  thermodynamics, the very large number of molecules in a macroscopic
  gas container which caused Maxwell to formulate his statistical
  kinetic theory, Boltzmann's statistical approach which attempted to
  reconcile irreversibility as manifest in the second law of
  thermodynamics with Newtonian mechanics, the discovery of
  radioactivity and of discrete spectral lines in the emision of light
  from stars and atoms that could not be understood in classical
  terms, Poincare's discovery of instable orbits in the astronomical
  three-body problem, and black-body radiation that led Planck to the
  discovery of the quantum of action. A further cause was the need to
  compensate for incomplete knowledge. Wigner introduced
  random-matrix theory because the nuclear Hamiltonian was largely
  unknown. That theory now governs the statistical theory of nuclear
  reactions. It has spread to many fields of physics, from the theory
  of electron transport through mesoscopic samples to quantum
  chromodynamics, to Andreev scattering, to the SYK model used in
  astrophysics and condensed-matter theory, to two-dimensional
  gravity, to random quantum circuits, and even to number theory.

\end{abstract}
  
\maketitle

\section{Introduction}

On the occasion of the retirement of two distinguished members of the
Institute of Theoretical Physics of the University of Cologne, I was
invited to give a survey talk about the historical development of
modern views of randomness in physics. The present paper is the result
of that invitation. It presents a physicist's view of the development,
without claim to scientific rigor or completeness in the historian's
sense. The text comes with a set of references. These are restricted
to papers published after 1945. Earlier references are usually
difficult to come by and, in fact, seldom read. I have tried to select
those papers that shed light on my narrative and/or that I consider
most important or most characteristic of the historical
development. The choice is necessarily subjective.

In the historical development of the physical sciences, statistical
concepts have played a significant role only since the middle of the
nineteenth century. They were introduced by Maxwell (1831 - 1879),
Gibbs (1839 - 1903), and Boltzmann (1844 - 1906). But in fields
different from physics, such concepts were introduced much
earlier. They actually had a precursor already in classical antiquity,
in politics, in philosophy, and in religion. I begin with a brief
account of these developments.

\section{From Classical Antiquity to Modern Times}

In classical greek democracy, chance played a big role. In Athens,
every male adult citizen could take part in the administration of the
city, had the right to vote, and was eligible for most offices. Women,
slaves, and non-citizens were excluded. The total number of voters at
the time of Perikles (about 490 to 429 B.C.) is estimated to have been
about 30 000 to 35 000. There were no professional judges nor
professional city administrators. All 6000 judges and most of the 700
city administrators were drawn periodically by lot from the population
of male citizens older than 30 years. To that end the Athenians used a
randomization device (the kleroterion) filled with white and black
spheres. A white sphere indicated that a person or group of persons
was elected, a black sphere the opposite. That stochastic procedure
was in use for several hundred years. It guaranteed broad
participation of all parts of (male) society in the affairs of the
city~\cite{Geb24}.

Chance played a role also in the natural philosophy of ancient Greece.
In his book ``De rerum natura'' the Roman author Titus Lucretius Carus
(approximately 99 - 53 B.C.) describes the natural philosophy of the
Greek philosopher Epicurus (about 340 to 270 B.C.)~\cite{Luc14}. That
philosophy was built upon ideas of Democritus (about 450 to 370
B.C.). All processes in nature are the result of the motion of
unchangeable and undivisible elements called atoms. These follow
deterministic laws of motion. Complete determinism is avoided by
postulating that without any external cause, the atoms occasionally
deviate from the prescribed trajectories and make a swerve on their
own. That hypothesis introduces a stochastic element. Epicureanism
became obsolete already during late Roman times and was totally
forgotten eventually. But the book of Lucretius was
preserved. Originally the book must have been written on papyrus. But
papyrus does not last fifteen hundred years (except under very special
circumstances). Some unknown monk librarian in some monastery must
have held the book in such high esteem that he had it copied onto
parchment, an expensive procedure (about one sheep per page) - and
that in spite of the fact that the book is purely materialistic and
denies the influence of the gods on human affairs. The book was
rediscovered by the humanist Poggio Bracciolini. Bracciolini was
secretary to Pope John XXIII and accompanied the Pope to the Council
of Constance (1414 to 1418). The Council had been summoned to resolve
the great schism that had led to three Popes claiming simultaneously
sovereignty over the church. The Council forced the three to abdicate
or it disposed of them, and Bracciolini lost his boss and his
employment. Being a humanist, he used his time to search for
manuscripts dating from classical antiquity in the nearby monasteries
in Southern Germany. He discovered the book by Lucretius (probably in
the monastery of Fulda). The book was multiply copied soon thereafter
and since then has had a strong influence on the intellectual history
of Europe. The story is described in detail in the book ``The Swerve''
by Stephen Greenblatt~\cite{Gre11}. It does not seem, however,
that the book by Lucretius caused or influenced the development of
stochastic concepts in the natural sciences.

Fortuna, the Roman goddess of fate and good fortune, was apparently
venerated from very early times. Numerous temples of the Roman World
were dedicated to her. Statues show her with her main attributes, the
wheel and the cornucopia. The wheel symbolizes rise and decline of a
person's life. The cornucopia was filled with flowers and fruit,
symbolizing fertility and wealth. It seems that with the collapse of
the western part of the Roman Empire, Fortuna was forgotten. But
beginning with the twelfth century, she was rediscovered. She was
considered a servant of God and played a role in monastic schools and
at the Universities, even though that role was actually incommensurate
with Christian faith. Illustrations of Fortuna carrying the wheel and
the cornucopia may be found in many books dating from the middle
ages. Her role in antiquity as goddess of fertility lost importance.
She was considered multi-faceted and fickle, distributing from the
cornucopia good and bad luck irrespective of a person's social
standing. Fortuna is popular even now. In 1803, a collection of
medieval songs and texts was discovered in the monastery
Benediktbeuren. Some of these address Fortuna. Carl Orff used these
texts in his ``Carmina Burana'', written in 1935/6. He dedicated parts
of this work to Fortuna. And many soccer clubs in Germany carry her
name.

The study of probability and statistics as mathematical disciplines
was triggered by gambling and by insurance. Insurance of shipments of
goods is related to gambling: Both activities involve a bet on the
outcome of an event that is beyond the control of either
party. Gambling is an old activity of mankind~\cite{Bri24}, documented
in ancient China, Rome, and Egypt. In the Bible it is reported that
the Roman guards cast lots over the garments of Jesus. In the
fifteenth century, organized gambling was introduced in the form of
lotteries aimed at raising money. Gambling houses were opened in the
seventeenth century. Insurance was triggered by trade. Beginning with
the early Renaissance, i.e., with the fourteenth century, wool, cloth
and other goods were transported in increasing quantities and over
large distances throughout Europe. Wool largely came from England,
cloth and silk from the Italian cities, spices from the Orient via
Venice. Silver and Copper from mines in Central Europe were added to
the list in the late fifteenth century. These goods were traded first
in the fairs of the Champagne, later in Bruge, Antwerp, and other
important towns. The trade was in the hands of a few families (the
Medici in Florence and the Fugger in Augsburg, for instance). The
Fugger became fabulously rich. That fact is indicative of the volume
of trade through Europe. But the roads were insecure, and shipwreck
always loomed. It was possible to insure shipments. Insurance is
mentioned in several biographies of leading Renaissance merchants by
Wescher~\cite{Wes35}. And Alberto Tenenti characterizes insurances as
a ``fundamentally new element ... in the mode of life ... of the
renaissance merchants"~\cite{Ten90}. Insurance rates were purely
empirical for a long time but obviously needed a more secure
foundation. In the 17$^{\rm th}$ century, a gambling problem caused
Blaise Pascal and Pierre Fermat to develop a mathematical approach
that led to probability theory. Probability theory never became a
fully independent mathematical discipline, however. Its development
remained intertwined with its applications. Starting with gambling and
insurance, these extended over an ever growing range of the natural
and social sciences~\cite{Gig89}. Curiously, probability theory
did not reach the core of the physical sciences until the middle of
the nineteenth century.

\section{The Mechanical world view and its challenges}

That is because in physics, Newton's universal theory of motion,
developed in the 17$^{th}$ century and made into an imposing edifice
by Euler, Laplace, Hamilton and others, had had overwhelming
success. In the eyes of its founders, all motion in nature follows
Newton's laws. These laws, not probability, ruled nature. There was no
room for probabilistic elements. The equations of the theory contain
the forces of gravitation and perhaps other, non-gravitational forces
like elasticity, and are supplemented by a set of initial conditions
at some fixed time. That theory had led to a unified understanding of
the motion of the celestial bodies and of bodies on earth, including
the tides of the oceans, the motion of thrown and falling bodies, the
flight course of cannon balls, etc. It is difficult for us today to
appreciate the enormous impact of that universal theory of motion on
the contemporaries of Newton and on the generations to follow. The
success of the theory led physicists to adopt what became known as the
mechanical world view. According to that view, all physical processes
can be understood on the basis of Newton's equations. The theory is
completely deterministic. There is no room whatsoever for
probabilistic concepts which play a role only in the analysis of
statistical and systematic errors. Such errors were considered
epistemic and, therefore, did not challenge the validity of the
mechanical world view. (A notable exception is the discovery of the
dwarf planet Ceres in 1801. After its discovery, the planet was lost
and could be retraced only with the help of Gauss' statistical
least-squares method). With the discovery of Maxwell's equations in
the 1860's, it may have seemed that a complete description of the
physical world had been accomplished. The remaining discrepancy
between Newton's mechanics (which is invariant under Galilean
transformations) and Maxwell's equations (which are invariant under
Lorentz transformations) was removed when Einstein introduced special
relativity in 1905.

Actually, however, that view was challenged almost as soon as the
apparently all-encompassing deterministic theory comprising mechanics
and electromagnetic theory had been set up. Serious problems with the
mechanical world view began to emerge in the middle of the nineteenth
century. Each of the following discoveries posed a separate challenge
to that view. In 1850, Clausius had formulated the second law of
thermodynamics. In 1865, Loschmidt had determined the number of
molecules of an ideal gas in a given volume (actually, the existence
of molecules had not been definitely demonstrated yet. That happened
only in 1905 when Einstein identified Brownian motion as a fluctuation
phenomenon due to molecular collisions). In 1896, Bequerel had found
that Uranium compounds emit radiation spontaneously without any prior
course. The spectral analysis of Bunsen and Kirchhoff developed in
1859 showed that light emitted from atoms and from the stars came in
discrete frequencies that were specific to a chemical element. And in
the 1890's, Poincare had shown that the three-body problem in
astronomy possesses instable orbits. Each of these developments
challenged the mechanical world view in a very different way and
triggered completely new developments. Some of the problems raised
turned out to be unsurmountable within Newtonian mechanics. The
problems were:

(i) Loschmidt's number. Even if the roughly 10$^{23}$ gas molecules in
a macroscopic container were to obey Newton's equations with known
forces describing their mutual interaction and their interaction with
the walls of the container, it is obviously out of the question to
determine at some fixed time the initial conditions for each molecule,
not to speak of solving the resulting differential equations. That
problem did not challenge the mechanical world view as such. It may
simply be a problem of large numbers. It shows, nevertheless, that the
mechanical world view cannot provide a complete description of the
physical world.

(ii) Irreversibility. The second law of thermodynamics is at variance
with Newton's equations. These second-order differential equations in
time are invariant under time reversal. If Newton's equations were to
rule the physical world, the inverse in time of every type of motion
that actually exists should likewise occur. That contradicts daily
experience and the second law. The problem was how to reconcile the
mechanical world view with irreversibility.

(iii) Radioactive Decay. The spontaneous radioactive decay of Uranium
and some other heavy elements was experimentally shown to follow a
purely statistical law, an exponential function characterized by a
constant, the mean life time. It became evident that such a completely
random process cannot be accounted for in the framework of Newton's
equations.

(iv) The discrete spectral lines of light emitted from atoms and stars
did not call with the same urgency for a statistical interpretation as
did radioactive decay. But the phenomenon as such had no explanation
within classical physics and challenged the mechanical world view.

(v) Chaotic Motion. Poincare wrote ``It may happen that small
differences in initial conditions cause big differences in the
results. Then a prediction becomes impossible, and we deal with a
random phenomenon''. That was the discovery of deterministic
chaos. The resulting limit on the predictability of mechanical systems
also limits the validity of the mechanical world view.

Actually problems (iii) and (iv) showed a fundamental shortcoming of
the mechanical world view. That view had been restricted to the
deterministic description of the motion of physical bodies. The
existence of such bodies is a basic assumption of the theory. It is
not an explanandum. The structure of these bodies (more generally and
in modern parlance: the structure of matter) was beyond the scope of
the mechanical world view and was addressed, if at all, by
chemists. Problems (iii) and (iv) showed that understanding the
structure of matter was a problem of physics just as much as was the
understanding of the motion of bodies. That opened the door to an
entirely new range of topics for physical investigations. Relevant
questions were: What causes the atoms of a given element to be
absolutely identical? What causes them to emit light in specific
frequencies only? What causes atoms to form stable molecules? What
causes molecules to form stable solids? What causes radioactivity? It
turned out that these questions can be answered only in the framework
of quantum theory, a fundamentally non-deterministic statistical
theory.

In the course of time, each of the problems listed under (i) to (v)
was addressed with the help of statistical or probabilistic
concepts. Some of these were consistent with the mechanical world
view, others required a fundamentally different approach. But for all
these problems, probability has played a central role. Thus, these
five problems mark the beginning of the reign of probability in
physics.

Problem~(i) (the problem of large numbers) was adressed by Maxwell. In
three papers published in 1860, 1865 and 1878 (the first prior to the
discovery of Loschmidt's number), he developed the kinetic theory of
gases in thermal equilibrium, assuming that a gas consists of a large
number of molecules that keep colliding with each other and with the
walls of the container. That assumption forced Maxwell to replace the
solutions of Newton's equations by a statistical approach. It was the
first time in the history of physics that a statistical approach was
employed for the description of an actual physical system. The result
was the Maxwell-Boltzmann distribution (named also after Boltzmann
because of Boltzmann's contributions to problem (ii) addressed below).
Maxwell's approach is consistent with and actually complements the
mechanical world view. The statistical description may be seen as
mirroring our incomplete knowledge of the system at hand. In his 1878
paper, Maxwell took an important further step by introducing the
concept of the ensemble. The ensemble consists of a ficticious set of
physical systems that agree with the given one in observable
macroscopic properties but differ in microscopic specifications. The
concept was enlarged by Gibbs. The use of ensembles has become a
standard tool in statistical mechanics. It often simplifies the
calculation because ensemble averages are usually easier to do than
averages over the actual physical system at hand. Newtonian mechanics
predicts the motion of every constituent body and, thus, gives a
complete description of a dynamical system. The use of ensembles
abandons that description. It is confined to the calculation of
averages and, at best, fluctuations about the mean values. But the use
of ensembles is not in obvious disagreement with Newtonian mechanics.

Problem~(ii) (irreversibility) was addressed by Boltzmann. To account
for the second law in mechanical terms, Boltzmann in 1872 went beyond
Maxwell and considered not only equilibrium states but all states of a
physical system. Like Maxwell, Boltzmann used statistical concepts but
supplemented these by a counting procedure. He characterized every
macroscopic state of a mechanical system by the number of its
microscopic realizations. The most likely macroscopic state is the one
with the maximum number of such realizations. Boltzmann postulated
that the intrinsic motion of each mechanical system leads for purely
statistical reasons from a given initial macroscopic state necessarily
to the most likely one. That process is irreversible in time and
accounts for the second law. The approach to equilibrium is described
by an equation that carries Boltzmann's name. Boltzmann also gave an
explicit expression for the entropy.

Boltzmann's approach raises a number of questions that have been
and/or are still being argued about. Does Boltzmann's basic assumption
(the system's motion always leads to the most likely macroscopic
state) follow from Newton's law? Or is it an additional postulate? If
so, is Boltzmann's approach consistent with Newton's law? Loschmidt
argued that reversing the motion of all constituents of the mechanical
system at a given time would lead to a decrease in entropy (which
contradicted the second law). Boltzmann answered with a probabilistic
argument: Such motion is possible but is so unlikely that it will
never be observed. The controversy established the second law as a
probabilistic law, in contrast to the first law which is strict. For
the first time in history, a fundamental law of physics was understood
to hold on average only. In 1867, Maxwell arrived at the same
conclusion with the help of his demon. The second law may be violated
for sufficiently small physical systems for which fluctuations about
the thermodynamic mean values become sufficiently large. Modern
experiments on sufficiently small systems have confirmed that
view, see~\cite{Jar11}.

Boltzmann's work and work by his followers, both in classical and in
quantum statistical mechanics, has led many physicists to the opinion
that thermodynamics is no more than an appendix to statistical
mechanics. In graduate courses on statistical mechanics, the three
laws of thermodynamics are often presented as direct consequences of
the statistical theory, a detailed presentation of the conceptual
approach used in thermodynamics is suppressed. I strongly feel that
this view does not do justice to the edifice of thermodynamics, built
in the $19^{\rm th}$ century by engineers upon strictly logical
foundations and, in fact, a physical theory in its own right, free of
the problems that the statistical approach entails.

Problem~(iii) (radioactive decay) and problem~(iv) (spectral lines)
were the first quantum phenomena discovered experimentally. But the
quantum world was actually discovered in the context of black-body
radiation by Max Planck, not in the study of radioactive decay or of
light emission from atoms, and I now recall that
story~\cite{Jos95}. As a professor in Kiel and later in Berlin, Planck
had worked on thermodynamic problems all his professional life. As
remarked above, since its inception thermodynamics had been a field of
physics of its own, governed by the first and second law (the third
law was discovered only around 1906), and free from probabilistic
elements. For Planck, the second law was as strict a law of nature as
the first law. When confronted with Maxwell's and Boltzmann's
interpretion of the second law as a statistical one that might be
violated in special cases, Planck was appalled. He viewed Boltzmann's
result as due to the counting procedure and to the underlying
assumption that mechanical systems consist of discrete elements, i.e.,
atoms or molecules. (I recall that the existence of atoms and
molecules had not been confirmed yet experimentally.) As a
counterexample to Boltzmann's case, Planck, therefore, looked for a
system that was completely continuous, i.e., did not have discrete
atomic structure. The electromagnetic field seemed a viable
candidate. Thus, beginning in 1894, Planck turned to black-body
radiation. Boltzmann showed in 1897 that Maxwell's equations are
time-reversal invariant. He was, therefore, convinced of the futility
of Planck's attempt to derive irreversibilty in that way and
formulated his critique in several publications. That did not deter
Planck.

The spectrum of black-body radiation in dependence on the temperature
of the emitting black body had been measured by Rubens and Kurlbaum at
the Physikalisch-Technische Reichsanstalt in Berlin. (Their setup can
still be seen there today in what is now called the
Physikalisch-Technische Bundesanstalt). Planck set out to derive the
spectrum from thermodynamics. At first, the formula he obtained seemed
to agree with the data. But more precise measurements revealed a
discrepancy. Leaving thermodynamics aside, in the year 1900 Planck
found a new formula by interpolating between known expressions for low
and high optical frequencies. That formula agreed very well with the
data. But how to justify it from first principles? Planck
unsuccessfully tried many avenues. He writes that the weeks following
the discovery of his successful interpolation formula were filled with
the hardest work of his life. All his attempts failed. He writes: ``In
my despair, I turned to the method Boltzmann''. Contrary to all his
long-held scientific convictions and irrespective of their earlier
dispute, he accepted Boltzmann's approach and postulated that the
electromagnetic field consists of discrete units of energy. The
discretization allowed Planck to apply Boltzmann's counting
procedure. That gave him the black-body radiation law which now
carries his name. The relentless pursuit of a problem Planck
considered important, combined with his ability to perform a complete
turnaround in his thinking and to dispense with life-long held deep
scientific convictions in favour of an approach that did account for
the facts, shows Planck's scientific genius. Planck's black-body
radiation law carries two constants. One is the quantum of action $h$
now called Planck's constant. In the context of Planck's radiation
law, $h$ describes the quantization of electromagnetic energy in units
of $h \nu$ where $\nu$ is the frequency. Since Einstein's 1905 paper
on light quanta, Planck's constant governs all of quantum physics. The
second constant is $k$. It relates temperature to energy. Like $h$,
that constant was discovered by Planck. But Planck named it after
Boltzmann in recognition of the latter's work even though the
``Boltzmann constant'' $k$ does not ever occur in Boltzmann's
writings. It is a curious fact that Boltzmann's gravestone in Vienna
carries Planck's formula for the entropy, $S = k \log W$.

The introduction of Planck's constant $h$ marked the beginning of the
quantum revolution in Physics. Planck had no further part in that. In
his work published after the discovery of the quantum of action, he
never tried to probe more deeply into the origins and causes of his
discretization approach. It was Einstein who realized that Planck had
actually counted quantum states of the photon. But the present account
focuses on the history of probability, not of quantum theory. Suffice
it to say, therefore, that the evolution of classical quantum theory
came to an end with Born's work in 1926 and the interpretation of the
square of Schr{\"o}dinger's wave function as a probability
density. Since then it has been accepted that quantum theory is an
intrinsically probabilistic theory. The time evolution of probability
amplitudes is deterministic, the results of experiments are not. That
closes the circle to Bequerel's data on Uranium decay. The decay of
each radioactive nucleus is a truly random process, unpredictable
individually but for a large set of nuclei firmly governed by a
statistical law.

Even today some of us find it difficult to accept that the laws of
nature should contain probabilistic elements. Over the years, several
serious attempts have been made to derive the quantum laws from an
underlying non-statistical theory. The experimental confirmation of
Bell's inequalities~\cite{Pan00} has put a final stop sign on all such
attempts.

Problem~(iv) (chaotic motion) was for a long time considered a special
problem in planetary motion and did not receive much attention from
the general physics community, for two reasons. First, there was no
evidence for chaotic motion in the solar system. Newtons' equations of
motion, later corrected for relativistic effects, adequately accounted
for all astronomical observations and made correct predictions.
Second, the study of mechanical systems by theoretical physicists had
for centuries been confined to the construction of analytical
solutions, later occasionally augmented by painfully slow numerical
calculations on the mechanical desk computer. That must have given the
impression that regular (i.e., non-chaotic) motion was all there was
to Newton's theory. Following the lead by Poincare, mathematicians
penetrated the problem more deeply. The analytical approach to
classical chaos was developed particularly by Kolmogorov in the
framework of ergodic theory in the 1940's and 1950's. Many basic
concepts of chaos theory were developed that shaped the field. Chaos
did not become a broadly accepted topic of physics, however, until the
late 1970's. Chirikov proposed a criterion for the emergence of
classical chaos in Hamiltonian systems. He also addressed chaos in
quantum systems. His work~\cite{Chi79} might not have received the
strong echo it did except for the development and general availability
of the electronic computer. With the help of that device it became
possible for every physicist to convince himself in simple ways of the
existence of instable orbits in mechanical systems and of their
consequences. Chaos was further popularized by the butterfly effect
discovered by Lorenz.

Instability under changes of the initial conditions manifests itself
in the exponential divergence of trajectories originating in
neighboring points of phase space. The exponential function carries
the leading Lyapunov coefficient. Instability is a feature of many (if
not most) Hamiltonian systems. Initial conditions cannot ever be
determined exactly. Solutions for the equations of motion can be found
only numerically. Rounding errors are unavoidable and cause the
numerical solutions to differ qualitatively from the exact ones. As
clearly perceived by Poincare, the inherent instability of chaotic
classical dynamical systems makes it impossible to predict the orbits
beyond a characteristic time scale given by the leading Lyapunov
coefficient. Beyond that scale, only a probabilistic description of
the system is possible.

Independent of these developments but in line with the concepts
developed by Maxwell and Boltzmann, Holtsmark~\cite{Hol19} in 1919
used a statistical approach to describe the fluctuating
electromagnetic fields in a plasma that are due to the random motion
of charged particles. In the 1940's, Chandrasekhar and von Neumann
used that approach to model fluctuating gravitational fields due to
the random motion of gravitational bodies. The Holtsmark distribution
plays an important role in plasma physics and in astrophysics. 

\section{Random matrices}

To the four probabilistic approaches to physics considered so far,
related to Loschmidt's number, to irreversibility, to quantum theory,
and to chaos, a fifth one was added after world war two by Eugene
Wigner. Wigner had worked on nuclear reactors during the war. He was
very well acquainted with Bohr's concept of the compound nucleus. At
that time, the compound nucleus was essentially a black box. It was
considered a liquid consisting of a large number of strongly
interacting particles. After the war Wigner turned to the theory of
nuclear reactions. Virtually nothing was known about such reactions
except that they proceeded via the formation of the compound
nucleus. Therefore, Wigner was forced to formulate a very formal
theory. In all break-up channels (open or closed), fictitious boundary
conditions were imposed. The resulting Hermitean Hamiltonian enclosed
by the said boundaries yields a set of eigenfunctions and eigenvalues.
Coupling the Hamiltonian via Green's theorem to the actual physical
channels, Wigner and Eisenbud~\cite{Wig47} expressed nuclear reaction
cross sections in terms of the eigenvalues and the projections of the
eigenfunctions onto the channel wave functions of that Hamiltonian. In
the framework of the one- or few-level approximation, that theory is
used to analyse data on nuclear resonance reactions still today.

To fill the resulting formal structure with life, it was neccessary to
address the question: What can be said about the Hamiltonian of a
system if nothing is known about the forces between the constituents
except that they are strong, have short range, and the system is,
therefore, complex, i.e., has compound-nucleus features? (Here, I
follow the narrative of Bohigas and Weidenm{\"u}ller~\cite{Boh15}).
Thinking about the question Wigner discovered by accident in 1954 a
book by the mathematician Wilks that contained an account of the
Wishart ensemble of random matrices. He immediately realized that that
was the answer to his question. He first adopted the Wishart ensemble
but soon switched to the Gaussian ensemble. That gave birth to the use
of random matrices in physics. In the following years, attention was
focused mainly on establishing general properties of random-matrix
ensembles. The compound nucleus stepped into the background. Dyson
showed that there exist three random-matrix ensembles, the unitary,
the orthogonal, and the symplectic ensemble which are either complex
Hermitean, real symmetric, or quaternion self-dual. Porter, Thomas,
Mehta, Rosenzweig and Gaudin derived important statistical properties
of the eigenfunctions and eigenvalues and/or introduced non-invariant
generalizations of the ensembles. These developments are summarized in
the books by Porter~\cite{Por65} and Mehta~\cite{Meh04}. As a
statistical theory that uses the concept of ensembles, random-matrix
theory obviously cannot replace the actual Hamiltonian of a system. It
does not specify quantitatively the eigenvalues and eigenfunctions.
It makes statements about the probability distribution of these
quantities. These can be tested against data.

In view of their later success and importance, it is astonishing that
random matrices remained dormant in physics for a number of years. In
nuclear physics, the field of their birth, the enormous success of the
nuclear shell model (discovered independently in 1949 by
Goeppert-Mayer~\cite{Goe49} and by Haxel, Jensen, and
Suess~\cite{Hax49}) and of the collective model (discovered by Bohr
and Mottelson in 1953~\cite{Boh53}) totally overshadowed the idea of
the compound nucleus and a probabilistic approach towards it. Both
models are fully deterministic. Attention of the overwhelming majority
of nuclear physicists turned to the experimental and theoretical
exploration of the low-lying parts of nuclear spectra and associated
transition probabilities across the chart of nuclides. The
experimental and theoretical study of nuclear resonance reactions
mainly served the purpose of furnishing spectroscopic data. In
addition, it turned out to be very difficult to obtain clear
experimental spectroscopic evidence for the validity of random-matrix
predictions for nuclei. That required experimental data on large,
pure, and complete sequences of nuclear states (i.e., sequences with
unambiguous identification of energy, spin, and parity of all levels
and the certainty that no levels were missed). In 1958/9 I spent a
year as postdoc at the University of Minnesota where I met Charles
Porter. Porter had taken a year's leave of absence from Brookhaven
National Lab in the hope of becoming appointed there. He was
disappointed by the lack of an echo the work on random matrices had
found in the nuclear-physics community. That lack of an echo was
probably the reason why Porter was not offered the professorship he
had hoped for and eventually had to return to Brookhaven. It was only
in 1982 that validation of random-matrix theory in terms of nuclear
spectral properties was announced by Haq, Pandey, and
Bohigas~\cite{Haq82}. Their claim has not remained
unchallenged~\cite{Koe10}. Shell-model calculations for nuclei in the
middle of the $s-d$-shell~\cite{Zel96} have shown that the residual
interaction of the shell model mixes the basic states of the
shell-model so strongly that distributions of the resulting
eigenvalues and eigenfunctions approach, for sufficiently large matrix
dimension, predictions of random-matrix theory. That insight bridges
the gap between a deterministic dynamical theory (the shell model) and
the statistical approach used for the compound nucleus.

During the last fifty years random matrices have found their way into
many fields of physics and of mathematics. That is impressively shown,
for instance, in the contributions to the ``Oxford Handbook on Random
Matrix Theory''~\cite{Oxf15}. In physics, that success is probably
largely due to the fact that random matrices are capable of simulating
complex systems, irrespective of any dynamical
details~\cite{Guh98}. In the sequel we highlight some of these
developments.

In spite of the overwhelming focus of the community on nuclear
spectroscopic properties and nuclear structure, nuclear-reaction
theory was pursued by several groups. Wigner's work on nuclear
reaction theory and on random matrices obviously called for a
combination of the two, i.e., for a statistical theory of nuclear
reactions. That is, after all, why Wigner introduced random matrices
in the first place. Hopefully, the theory would validate, lead to a
deeper understanding and, at the same time, also show the limitations
of Bohr's picture of the compound nucleus. Setting up such a theory
turned out to be difficult. Feshbach, Kerman and coworkers at MIT
addressed the problem. The Hauser-Feshbach formula~\cite{Hau52}
implements Bohr's postulate of the independence of formation and decay
of the compound nucleus but is purely penomenological and not based
upon a random-matrix approach. The general theory formulated by Kawai,
Kerman, and McVoy~\cite{Kaw73} uses ad-hoc statistical assumptions on
parameters that appear in the course of the derivation, without
connection to random-matrix theory. The most extensive effort on
deriving properties of nuclear cross sections from an underlying
statistical theory was undertaken by Moldauer~\cite{Mol75}. Moldauer
worked as a theorist in the reactor department of Argonne National Lab
and was confronted daily with the need to calculate reliably nuclear
reaction cross sections for medium-weight and heavy nuclei. He used a
representation of the scattering matrix developed by Humblet and
Rosenfeld~\cite{Hum61}. In the complex energy plane, each element of
the scattering matrix (an analytic function of energy) is written as
the sum of contributions from the pole terms plus a smooth
background. Each pole term represents a resonance (a level of the
compound nucleus) and is characterized by the location of the pole (a
complex energy) and by the residue (which defines the coupling of the
resonance to the channels). The approach has the advantage that the
poles are determined by intrinsic properties of the system. The
approach avoids arbitrary boundary conditions as used in Wigner's
nuclear-reaction theory. Moldauer made assumptions on the distribtion
of the pole parameters. In an impressive sequence of papers published
over a time span covering two decades until his death in 1984, he
obtained some lasting results but did not fully solve the problem. In
hindsight it appears that the Humblet-Rosenfeld theory is
fundamentally unsuited for a statistical approach. Except for the
special case of isolated resonances, unitarity of the scattering
matrix implies strong correlations between the pole parameters
pertaining to different resonances. It does not seem possible to
formulate statistical assumptions on these quantities that would
respect unitarity. Ericson~\cite{Eri60} also used the pole expansion
of the scattering matrix. The presentation he used did not respect
unitarity. Nevertheless, combining that presentation with statistical
assumptions on the resonance parameters he correctly predicted the
existence of random fluctuations versus energy and angle of compound
nuclear reaction cross sections. In the following years, the advent of
Van-de-Graaf accelerators of sufficient energy and with sufficient
energy resolution made it possible to verify the existence of these
fluctuations experimentally and to investigate their
properties~\cite{Eri66}. These developments redirected the attention
of the nuclear physics community towards statistical concepts. They
confirmed that such concepts must be an integral part of
compound-nucleus theory.

The approach to compound-nucleus reactions developed in
Heidelberg~\cite{Mah69} was inspired by the nuclear shell model. The
approach avoids Wigner's construction and the use of boundary
conditions. It also avoids the pole expansion used by Humblet and
Rosenfeld. It expresses the elements of the scattering matrix directly
in terms of the Hamiltonian of the bound system and its couplings to
the channels. The explicit occurrence of the Hamiltonian in the
propagator of the scattering matrix makes it possible to implement
statistical assumptions directly on the Hamiltonian, that is, to
replace the actual nuclear Hamiltonian by a random-matrix ensemble. It
became possible to calculate average cross sections by numerical
simulation and/or by asymptotic expansion. The missing last step, the
analytical calculation of average cross sections from random-matrix
theory~\cite{Ver85}, used the invariance property of the orthogonal
ensemble and Efetov's supersymmetry technique. The result defines the
range of validity of the compound-nucleus picture and its
limitations. It yields the Hauser-Feshbach formula as a limiting
case. The analytical expression for the average compound-nucleus cross
section given in Ref.~\cite{Ver85} has been since used as a bench mark
for approximations and has been implemented in numerical programs for
compound-nucleus reaction cross sections. Such calculations are needed
for nuclear reactions on unstable targets that cannot be measured in
the laboratory but do occur in practice and are important in reactor
physics, in astrophysics, in medicine, and elsewhere.

Similar developments took place in mesoscopic physics. Scattering
theory was used to describe the transport of electrons through a
disordered medium. The actual distribution of the impurities on which
the electrons are scattered is not known in detail. It is simulated by
an ensemble of randomly distributed impurities. Using that model,
Altshuler~\cite{Alt85} and Lee and Stone~\cite{Lee85} independently
worked out the theory of universal conductance fluctuations in
mesoscopic samples.

In nuclear physics, the use of random-matrix theory had encountered
critique because in the shell model, the residual interaction is
dominated by two-body interactions whereas random-matrix theory
implicitly assumes many-body forces. That caused French and
Wong~\cite{Fre70} and Bohigas and Flores~\cite{Boh71} to propose a
random ensemble of two-body interactions instead, to be added to the
mean field of the shell model. More generally, embedded ensembles with
$k$-body interactions with $k = 2, 3, \ldots$ were considered. Such
models are very difficult to handle analytically. A simplification and
extension is the Sachdev-Ye-Kitaev (SYK) model. It disregards the
mean-field altogether. Majorana Fermions are coupled to each other
with a random, Gaussian-distributed four-body interaction. That model
has played a big role in the physics of black holes and in
condensed-matter physics, see Ref.~\cite{Ros19} for a review.

In the 1990's, Verbaarschot found that the spectrum of the Dirac
operator in Quantum Chromodynamics at low energy coincides with a new
variant of random-matrix theory that he called chiral random-matrix
theory. He discovered three chiral ensembles of random matrices,
distinguished by their symmetry properties~\cite{Ver94}. The
discovery, first viewed with skepticism by the community, has been
fully accepted and has turned out to be extremely helpful for lattice
gauge calculations. It provides guidance in the extrapolation of
numerical results to infinite lattice size. Beyond that, it has been
shown that the connection is exact: the low-energy spectra of chiral
field theories exactly coincide with those of the corresponding chiral
random-matrix theory. In an independent development, the study of
Andreev scattering in condensed-matter physics in the 1990's revealed
the existence of four new random-matrix
ensembles~\cite{Alt97}. Andreev scattering is the process where an
electron propagating in a disordered mesoscopic sample is scattered on
the surface of a superconductor. These four new ensembles together
with the three chiral ensembles supplement the three classical
ensembles introduced by Dyson. Altland and Zirnbauer used group theory
to classify the totality of the altogether ten random-matrix ensembles
and to show that the list is complete~\cite{Alt97}.

A connection has also emerged between random-matrix theory and
two-dimensional quantum gravity. I am not sufficiently familiar with
the subject to do more than give a reference to chapter~30 in the
Oxford Handbook of Random Matrices~\cite{Oxf15}. Likewise, randomness
plays a big role in quantum circuits that provide a new arena for
quantum many-body physics and the possibility to explore universal
collective phenomena far from equilibrium~\cite{Fis22}.

I believe few people would have expected a connection to exist between
prime numbers and random-matrix theory. The connection emerged from
numerical studies of Riemann's Zeta function. The Zeta function is a
complex-valued analytical function defined in terms of the prime
numbers. In the complex plane, it possesses a number of trivial
zeros. The Riemann hypothesis says that the remaining non-trivial
zeros are all located on a straight line parallel to the imaginary
axis. Numerical calculations are in accord with the hypothesis. The
hypothesis is fundamental to an understanding of the distribution of
prime numbers. Proof of the hypothesis is considered one of the most
important open problems in pure mathematics. The connection to
random-matrix theory emerged when Odlytzko had calculated numerically
the first million or so prime numbers (later that number grew to about
70 million). Following a conjecture by Montgomery, he plotted the
distribution of the distances of nearest neighbors. In 1974,
Montgomery showed the plot to Dyson. Dyson realized that the plot was
similar to that of the nearest-neighbor spacing distribution of
eigenvalues of the Gaussian Unitary Ensemble (GUE) of random
matrices. Further numerical tests have confirmed the connection and
have shown that the distribution of non-trivial zeros of the Zeta
function seems to tend asymptotically towards that of the GUE. That
strongly suggests the existence of a complex Hermitean operator in
Hilbert space the eigenvalues of which coincide with the nontrivial
zeros of the Zeta function. For more information, see the review by
Keating and Snaith in chapter~24 of the Oxford Handbook of Random
Matrices~\cite{Oxf15}.

In summary, in the last 175 years physicists have been led or been
forced to ascribe an ever increasing role to probability in the
description of nature. I have listed four causes for that development:
Loschmidt's number supporting Maxwell's theory, irreversibility
leading to Boltzmann's approach, Bequerel's discovery, spectral lines
and black-body radiation leading to quantum theory, and Poincare's
discovery of classical chaos. Random-matrix theory is different. It
was not imposed on physicists by an experimental or theoretical
discovery but was introduced to compensate for the incomplete
knowledge of the Hamiltonian.

Are these probabilistic descriptions related or dependent upon each
other? Surely the probabilistic nature of quantum theory is in a class
of its own. But there may be a connection between classical chaos and
classical statistical mechanics, and between quantum chaos and
random-matrix theory. As for the first, I am not sufficiently familiar
with ergodic theory to make substantial statements. I refer the reader
to the book by Gaspar~\cite{Gas98} where further references may be
found. As for the second, Bohigas, Giannoni, and Schmit~\cite{Boh84}
have conjectured that upon quantizaton, the fluctuation properties of
the spectra of classically chaotic systems coincide with those of the
random-matrix ensemble in the same symmetry class. The conjecture has
been confirmed numerically for many systems. With the help of
Gutzwiller's periodic orbit expansion for the density matrix of a
chaotic system, the validity of the conjecture has been demonstrated
for the two-level correlation function. That is summarized in
chapter~30 of The Oxford Handbook~\cite{Oxf15}. The converse of the
Bohigas-Giannoni-Schmit conjecture is not true, however. There are
complex quantum systems that follow random-matrix predictions but do
not possess a classical limit. That is why random-matrix theory
probably goes beyond quantum chaos.

{\bf Acknowledgement.} I am grateful to Alexander Altland for
providing the initial momentum for this investigation and for useful
suggestions, to Eric Lutz for a stimulating discussion, and to him and
to Adriana Palffy for comments.

\end{document}